\documentclass[aps, prd, reprint, amsmath, amssymb, superscriptaddress, nofootinbib, showpacs]{revtex4}

\usepackage{graphicx}
\usepackage{dcolumn}
\usepackage{bm}
\usepackage{hyperref}
\hypersetup{
    colorlinks=true,       
    linkcolor=blue,        
    citecolor=blue,        
    urlcolor=blue           
}

\newcommand{\sms}{\scriptscriptstyle}

\begin{document}

\title{Effects of structure formation on the expansion rate of the Universe: An estimate from numerical simulations}
\author{Xinghai Zhao}
\email{xzhao@nd.edu}
\affiliation{Department of Physics, University of Notre Dame, Notre Dame, Indiana 46556, USA}
\affiliation{Center for Astrophysics, University of Notre Dame, Notre Dame, Indiana 46556, USA}
\affiliation{The Joint Institute for Nuclear Astrophysics, University of Notre Dame, Notre Dame, Indiana 46556, USA}
\author{Grant J. Mathews}
\email{gmathews@nd.edu}
\affiliation{Department of Physics, University of Notre Dame, Notre Dame, Indiana 46556, USA}
\affiliation{Center for Astrophysics, University of Notre Dame, Notre Dame, Indiana 46556, USA}


\begin{abstract}
General relativistic corrections to the expansion rate of the Universe arise when the Einstein equations are averaged over a spatial volume in a locally inhomogeneous cosmology. It has been suggested that they may contribute to the observed cosmic acceleration. In this paper, we propose a new scheme that utilizes numerical simulations to make a realistic estimate of the magnitude of these corrections for general inhomogeneities in (3+1) spacetime. We then quantitatively calculate the volume averaged expansion rate using N-body large-scale structure simulations and compare it with the expansion rate in a standard FRW cosmology. We find that in the weak gravitational field limit, the converged corrections are slightly larger than the previous claimed $10^{-5}$ level, but not large enough nor even of the correct sign to drive the current cosmic acceleration. Nevertheless, the question of whether the cumulative effect can significantly change the expansion history of the Universe needs to be further investigated with strong-field relativity.
\end{abstract}


\maketitle

\section{Introduction}
One of the most puzzling questions in modern cosmology is the nature and origin of the dark energy that is responsible for the present cosmic acceleration. Over the last decade, evidence has been accumulating from the type Ia supernova luminosity distance-redshift relation \cite{Riess:1998,Riess:2004,Perlmutter:1999} and other observations \cite{Fosalba:2003,Tegmark:2004,Eisenstein:2005,Hoekstra:2006,Spergel:2007} indicating that the Universe is accelerating. In the standard Friedmann-Robertson-Walker (FRW) cosmology model, this can be explained by introducing a dark energy term in the Friedmann equations. Observations indicate that the dark energy comprises more than 70\% of the mass-energy in the Universe. Numerous explanations have been proposed for the origin of the dark energy. The simplest explanation is that of a mass-energy that violates the strong energy condition either in the form of a cosmological constant \cite{Riess:1998,Riess:2004,Perlmutter:1999} or a quintessence \cite{Peebles:1988,Caldwell:1998}. However, this solution has unnatural fine tuning and coincidence problems. Alternately, one could assume that GR is not the complete theory on cosmological scales (such as in the Dvali-Gabadadze-Porrati ($DGP$) \cite{Dvali:2000} or $f(R)$ \cite{Capozziello:2003,Carroll:2004} models), however, no corroborating evidence for deviations from standard GR has yet been found. An alternative explanation of interest to the present work is that corrections to a FRW cosmology, due to the presence of local inhomogeneities, may introduce dark energy like terms in the equation of cosmic expansion.

While the Universe appears homogeneous and isotropic on large cosmological scales, this is not the case on smaller scales. Local inhomogeneity and structure always exist. It has been argued \cite{Ellis:1984,Ellis:1987,Ellis:2008}, therefore, that one should study the observational data in the context of a realistic lumpy universe first, instead of assuming that the FRW model is correct, to fit cosmological parameters. Hence, it is imperative to clarify how the inhomogeneities affect the interpretation of the observational data.

Even before the discovery of the cosmic acceleration, it was argued \cite{Dyer:1972} that when light propagates only through a nearly empty intergalactic medium, there is a dimming effect compared to the FRW model. This may affect the interpretation of the supernova luminosity distance-redshift data. Recently, one simple class of the inhomogeneous models, the Lema\^{i}tre-Tolman-Bondi (LTB) model, has been extensively studied \cite{Celerier:2000,Alnes:2006,Garfinkle:2006,Marra:2008}. In this model, it is assumed that we are living near the center of a spherical underdense region and the Universe may have one or many such regions. The supernova data can then be fit without introducing a dark energy term. Although we may indeed reside in an underdense region, the special symmetry in this model is not consistent with current observations of the large-scale structure of the Universe. Moreover, none of the realistic models of this class can fit all of the observed cosmological constrains \cite{Alnes:2006,Garfinkle:2006,Vanderveld:2006,Vanderveld:2008}.

While more general inhomogeneous cosmological models have been investigated using various methods \cite{Futamase:1988,Hui:1996,Seljak:1996a,Siegel:2005}, the domain averaging procedure proposed by Buchert \cite{Buchert:2000,Buchert:2001,Buchert:2003,Buchert:2008} has been of most interest recently. By using perturbation methods in this domain averaging procedure in a general synchronous gauge, Kolb et. al \cite{Kolb:2005,Kolb:2006} claimed that an effective negative pressure term can arise from averaging local fluctuations. However, it has been strongly argued \cite{Hirata:2005,Flanagan:2005,Ishibashi:2006,Vanderveld:2007,Behrend:2008,Paranjape:2008} that the effect from this procedure is either not a physical observable or not large enough to drive the current cosmic acceleration. Moreover, such perturbative analysis is not easily amenable to the nonlinear evolution of the large-scale structure. The lack of a realistic model and the limitation of the perturbation methods make it difficult to calculate the proposed effect accurately. To alleviate this problem, therefore, we describe here the first step toward the development of a relativistic numerical scheme to explicitly calculate the proposed effect of domain averaging. 

The paper is organized as follows: We first develop a theoretical scheme that enables us to calculate the magnitude of the proposed correction terms quantitatively using an N-body large-scale structure simulation as described in Sec. II. The details of the numerical simulations are discussed in Sec. III. The results are presented and discussed in Sec. IV followed by a summary in Sec. V. The main purpose of this paper is to present the formalism and to make initial numerical investigations of the effects from this domain averaging procedure on the expansion rate of the Universe. This study will shed light on whether this is a viable approach to explain the nature of the dark energy.

\section{General Relativistic corrections from domain averaging}

\subsection{Domain averaged expansion rate in an inhomogeneous cosmology}

Here, we will roughly follow the domain averaging procedure proposed by Buchert \cite{Buchert:2000,Buchert:2001,Buchert:2003,Buchert:2008}. Numerical calculations in general relativity are best formulated in the Arnowitt-Deser-Misner (ADM) formalism \cite{Arnowitt:1962,Smarr:1978,Wilson:2003}. Hence, we start with the general (3+1) ADM metric:
\begin{equation}
ds^2=-(\alpha^2-\beta_i\beta^i)dt^2+2\beta_idx^idt+\gamma_{ij}dx^idx^j ~~,
\end{equation}
where $\alpha$ is the lapse function denoting the lapse of proper time, $\beta_i$ is the shift vector describing the shift of coordinates, respectively, from one time slice to the next and $\gamma_{ij}$ is the spatial three-metric. 
The proper volume of an arbitrary domain $D$ in this scheme can then be defined as:
\begin{equation}
\label{vd}
V_{\sms D}=\int_{\sms D}\gamma\, d^3 x ~~,
\end{equation}
where $\gamma=\sqrt{det(\gamma_{ij})}$, and $det(\gamma_{ij})$ is the determinant of $\gamma_{ij}$.
We can then define the average of an arbitrary scalar field $\psi(\bm x,t)$ on the domain $D$ as:
\begin{equation}
\label{daverage}
\langle\psi(\bm x,t)\rangle_{\sms D}=\frac{1}{V_{\sms D}}\int_{\sms D}\psi(\bm x,t)\gamma\, d^3 x ~~.
\end{equation}
The time derivative of this domain average is then:
\begin{eqnarray}
\label{dpsiddt1}
\frac{\partial\langle\psi\rangle_{\sms D}}{\partial t}&=&\frac{\partial}{\partial t}(\frac{1}{V_{\sms D}}\int_{\sms D}\psi\gamma\, d^3 x) \nonumber \\
                                      &=&-\frac{\dot{V}_{\sms D}}{V_{\sms D}}\langle\psi\rangle_{\sms D}+\langle\dot{\psi}\rangle_{\sms D}+\frac{1}{V_{\sms D}}\int_{\sms D}\psi\dot{\gamma}\, d^3 x ~~.
\end{eqnarray}
Now, using the fact that for any invertible matrix A:
\begin{equation}
\frac{\partial\,det (A)}{\partial A_{ij}}=det(A)(A^{-1})_{ji} ~~,
\end{equation}
and the fact that $\gamma_{ij}$ is symmetric. The derivatives of $\gamma_{ij}$ can be written as:
\begin{equation}
\label{dgamma}
\frac{\partial\,\gamma^2}{\partial\,\gamma_{ij}}=\gamma^2\gamma^{ij}, \qquad \frac{1}{\gamma}\dot{\gamma}= \frac{1}{2}\dot{\gamma}_{ij}\gamma^{ij} ~~.
\end{equation}

We next choose the Eulerian gauge (shift vector $\beta_i=0$)\footnote {The purpose of this simplification and application of the conformally flat condition in Sec. IIB is solely to make easy use of current numerical large-scale structure simulation codes. The generalization of the expressions in Sec. IIA and IIB in the general ADM formalism is straightforward.} to reduce the ADM metric to:
\begin{equation}
ds^2 = -\alpha^2(\bm x,t)dt^2 + \gamma_{ij}(\bm x,t) dx^i dx^j ~~.
\end {equation}
In this gauge, the extrinsic curvature $K_{ij}$, which can be interpreted as the rate of change of the spatial metric $\gamma_{ij}$ along the normal vector, can be simply expressed as:
\begin{equation}
\label{kequation}
K_{ij}=-\frac{1}{2\alpha}\dot{\gamma}_{ij}, \qquad K=\gamma^{ij}K_{ij}=-\frac{1}{2\alpha}\dot{\gamma}_{ij}\gamma^{ij} ~~.
\end{equation}
By inserting the expression for $\frac{1}{2}\dot{\gamma}_{ij}\gamma^{ij}$ from Eq. (\ref{kequation}) into Eq. (\ref{dgamma}), we have:
\begin{equation}
\dot{\gamma}=-\alpha K \gamma ~~.
\end{equation}
The time evolution of the domain average in Eq. (\ref{dpsiddt1}), can then be expressed as:
\begin{equation}
\label{dpsiddt2}
\frac{\partial\langle\psi\rangle_{\sms D}}{\partial t}=-\frac{\dot{V}_{\sms D}}{V_{\sms D}}\langle\psi\rangle_{\sms D}+\langle\dot{\psi}\rangle_{\sms D}-\frac{1}{V_{\sms D}}\int_{\sms D}\alpha K \gamma \psi\, d^3 x ~~.
\end{equation}
The new dimensionless scale factor $a_{\sms D}$ for the domain $D$, and the new Hubble parameter $H_{\sms D}$ for this domain, can now be defined as:
\begin{equation}
\label{ad}
a_{\sms D} = \left(\frac{V_{\sms D}}{V_{{\sms D}_0}}\right)^{1/3}, \qquad H_{\sms D} = \frac{\dot{a}_{\sms D}}{a_{\sms D}} = \frac{1}{3}\frac{\dot{V}_{\sms D}}{V_{\sms D}} ~~,
\end{equation}
where $V_{\sms D}$ is the proper volume defined in Eq. (\ref{vd}) and $V_{{\sms D}_0}$ denotes the domain volume at the present time. $H_{\sms D}$ can be further expressed as:
\begin{equation}
\label{hd}
H_{\sms D}=\frac{1}{3V_{\sms D}}\int_{\sms D}\dot{\gamma}\, d^3 x=\frac{1}{3V_{\sms D}}\int_{\sms D}-\alpha K \gamma\, d^3 x=\frac{1}{3}\langle-\alpha K\rangle_{\sms D}=\frac{1}{3}\langle\Theta\rangle_{\sms D} ~~,
\end{equation}
where we introduce the trace of the expansion tensor $\Theta^i_j$ as $\Theta=\Theta^i_i=-\alpha K$. $H_{\sms D}$ is now the appropriate physical quantity to describe the overall expansion rate of the inhomogeneous domain $D$. In what follows, we calculate this quantity explicitly using numerical large-scale structure simulations as realistic representations of the evolution of the lumpy universe within the domain.

\subsection{Time evolution of the expansion rate and effective pressure}

We now derive the time evolution equation of the domain averaged expansion rate. Substituting Eq. (\ref{hd}) into Eq. (\ref{dpsiddt2}), we obtain an important commutation rule:
\begin{equation}
\frac{\partial\langle\psi\rangle_{\sms D}}{\partial t}-\langle\dot{\psi}\rangle_{\sms D}=\langle\Theta\psi\rangle_{\sms D}-\langle\Theta\rangle_{\sms D}\langle\psi\rangle_{\sms D} ~~,
\end{equation}
which for $\psi=\Theta$, gives:
\begin{equation}
\label{thetacommute}
\frac{\partial\langle\Theta\rangle_{\sms D}}{\partial t}-\langle\dot{\Theta}\rangle_{\sms D}=\langle\Theta^2\rangle_{\sms D}-\langle\Theta\rangle_{\sms D}^2 ~~.
\end{equation}
In the ADM formalism, the time evolution of the extrinsic curvature scalar $K$ can be written as \cite{Arnowitt:1962,Smarr:1978,Wilson:2003}:
\begin{equation}
\label{kdot}
\dot{K}=-D_iD^i\alpha+\alpha(^{3}R+K^2)+4\pi G\alpha(S-3\rho_H) ~~,
\end{equation}
where $D_i$ is the covariant derivative operator in the three-space which reduces to an ordinary gradient operator for a scalar such as $\alpha$. The quantity $^{3}R$ is the Ricci scalar for the three-metric $\gamma_{ij}$. $S=3P+\rho h(W^2-1)$ is the trace of the spatial stress, and $\rho_H=\rho hW^2-P$ is the Hamiltonian density. In the expressions above, $\rho$ is the rest mass-energy density, $P$ is the pressure, $W\equiv\sqrt{1+u_i u^i}$ is a generalized Lorentz factor, $u_i$ denotes spatial components of  the four velocity, $h=1+\epsilon+P/\rho$ is the specific enthalpy, and $\epsilon$ is the internal energy per unit mass.

The Hamiltonian constraint derives from the ``00" component of the Einstein equation. In the ADM formalism, it can be written as:
\begin{equation}
\label{hequation}
^{3}R+K^2-K_{ij}K^{ij}=16{\pi}G{\rho_H} ~~.
\end{equation}
In the slow-motion, low-temperature limit, i.e. $W \cong 1, ~ \epsilon \cong 0, ~ \rho_H \cong \rho, ~ S-3\rho_H \cong 3(P-\rho)$, Eq. (\ref{kdot}) then becomes:
\begin{equation}
\label{newkdot}
\dot{K}=-D_iD^i\alpha+\alpha(^{3}R+K^2)+12\pi G\alpha(P-\rho) ~~.
\end{equation}
Henceforth, we denote $^{3}R$ as $R$.

For the present application, there is no significant rotation, shear or gravity waves. Hence, we can adopt a conformally flat condition \cite{Wilson:2003} to further simplify the ADM metric to the form:
\begin{equation}
\label{ds}
ds^2=-\alpha^2(\bm x,t)dt^2+a^2_{\sms{FRW}}(t)\phi^4(\bm x,t)\delta_{ij}dx^idx^j~~,
\end{equation}
where $a_{\sms{FRW}}(t)$ is the scale factor in the FRW model limit which is generally different from $a_{\sms D}$ in Eq. (\ref{ad}), $\phi(\bm x,t)$ is the conformal factor denoting the local deviations from a homogeneous and isotropic curvature in the three-space. 
Now, the conformally flat three-metric $\gamma_{ij}$ takes the form:
\begin{equation}
\label{gammaij}
\gamma_{ij}=a^2_{\sms{FRW}}(t)\phi^4(\bm x,t)\delta_{ij}~~.
\end{equation}
In this formulation, $K_{ij}$ has diagonal elements only and $K_{ij}K^{ij}=-\frac{1}{3}K^2$, thus we have:
\begin{eqnarray}
\label{kequations}
K_{ij}=-\frac{1}{2\alpha}\dot{\gamma}_{ij}=-\frac{1}{\alpha}(\frac{\dot{a}_{\sms{FRW}}}{a_{\sms{FRW}}}+2\frac{\dot{\phi}}{\phi})\gamma_{ij} ~~, \nonumber\\
K=-\frac{3}{\alpha}(\frac{\dot{a}_{\sms{FRW}}}{a_{\sms{FRW}}}+2\frac{\dot{\phi}}{\phi}) ~~.
\end{eqnarray}
The Hamiltonian constraint can then be reduced to:
\begin{equation}
\label{reducedh}
R+\frac{2}{3}K^2=16{\pi}G\rho ~~.
\end{equation}
Using Eq. (\ref{reducedh}), the $\dot{K}$ equation [Eq. (\ref{newkdot})] can be rewritten as:
\begin{equation}
\dot{K}=-D_iD^i\alpha+\frac{1}{3}\alpha K^2+4\pi G\alpha(\rho+3P) ~~.
\end{equation}
Now employing the relation $\alpha \dot{K}=-\dot{\Theta}-\dot{\alpha} K$, we can express the domain averaged $\dot{K}$ equation as:
\begin{equation}
\label{alphakdot}
\langle\alpha\dot{K}\rangle_{\sms D}=-\langle\dot{\Theta}\rangle_{\sms D}-\langle\dot{\alpha} K\rangle_{\sms D}=-\langle\alpha D_iD^i\alpha\rangle_{\sms D}+\frac{1}{3}\langle\Theta^2\rangle_{\sms D}+4\pi G\langle\alpha^2(\rho+3P)\rangle_{\sms D} ~~.
\end{equation}
Using Eq. (\ref{thetacommute}), we can then rewrite Eq. (\ref{alphakdot}) as:
\begin{equation}
\label{thetadot}
\langle\Theta^2\rangle_{\sms D}-\langle\Theta\rangle_{\sms D}^2-\frac{\partial\langle\Theta\rangle_{\sms D}}{\partial t}=(\langle\dot{\alpha} K\rangle_{\sms D}-\langle\alpha D_iD^i\alpha\rangle_{\sms D})+\frac{1}{3}\langle\Theta^2\rangle_{\sms D}+4\pi G\langle\alpha^2(\rho+3P)\rangle_{\sms D} ~~.
\end{equation}
The $\partial\langle\Theta\rangle_{\sms D}/\partial t$ term can be rewritten as:
\begin{equation}
\frac{\partial\langle\Theta\rangle_{\sms D}}{\partial t}= 3\frac{\partial\left(\dot{a}_{\sms D}/a_{\sms D}\right)}{\partial t}=3\left(\frac{\ddot{a}_{\sms D}}{a_{\sms D}}-\left(\frac{\dot{a}_{\sms D}}{a_{\sms D}}\right)^2\right)=3\frac{\ddot{a}_{\sms D}}{a_{\sms D}}-\frac{1}{3}\langle\Theta\rangle_{\sms D}^2 ~~.
\end{equation}
Plugging this expression into Eq. (\ref{thetadot}), we have:
\begin{equation}
\label{qd}
3\frac{\ddot{a}_{\sms D}}{a_{\sms D}}+S_{\sms D}(\alpha)+4\pi G\langle\alpha^2(\rho+3P)\rangle_{\sms D}=\frac{2}{3}\left(\langle\Theta^2\rangle_{\sms D}-\langle\Theta\rangle_{\sms D}^2\right) ~~.
\end{equation}
Here, we denote $S_{\sms D}(\alpha) \equiv \langle\dot{\alpha} K\rangle_{\sms D}-\langle\alpha D_iD^i\alpha\rangle_{\sms D}$ because it behaves like a source term. The new term $\frac{2}{3}\left(\langle\Theta^2\rangle_{\sms D}-\langle\Theta\rangle_{\sms D}^2\right)$ is the kinematic backreaction term $Q_{\sms D}$ originally proposed by Buchert \cite{Buchert:2000,Buchert:2001}. 

Now, we can write the modified first Friedmann equation by introducing an effective energy density $\rho_{\rm eff}$,
\begin{equation}
\label{1friedmann}
H_{\sms D}^2=\left(\frac{\dot{a}_{\sms D}}{a_{\sms D}}\right)^2=\frac{8}{3}\pi G\rho_{\rm eff} ~~, \qquad \mbox{with} \quad \rho_{\rm eff}=\frac{1}{24\pi G}\langle\Theta\rangle_{\sms D}^2 ~~.
\end{equation}
Similarly, from Eq. (\ref{qd}), the modified second Friedmann equation can be written by introducing an effective pressure $P_{\rm eff}$,
\begin{equation}
\label{2friedmann}
\frac{\ddot{a}_{\sms D}}{a_{\sms D}} =  -\frac{4\pi G}{3}\left(\rho_{\rm eff}+ 3P_{\rm eff}\right) ~~, \qquad \mbox{with} \quad P_{\rm eff}=\frac{1}{3}\langle\alpha^2(\rho+3P)\rangle_{\sms D}+\frac{1}{12\pi G}S_{\sms D}(\alpha)+\frac{1}{24\pi G}\langle\Theta\rangle_{\sms D}^2-\frac{1}{18\pi G}\langle\Theta^2\rangle_{\sms D} ~~.
\end{equation}
Here, $\Theta=-\alpha K=3(H_{\sms{FRW}}+2\dot{\phi}/\phi)$ and $\Theta^2=\alpha^2 K^2=9(H_{\sms{FRW}}+2\dot{\phi}/\phi)^2$, where $H_{\sms{FRW}}$ is the Hubble parameter defined by the unperturbed Friedmann equation, $H_{\sms{FRW}}^2=({\dot{a}_{\sms{FRW}}}/{a_{\sms{FRW}}})^2=\frac{8}{3}\pi G\rho_{\sms{FRW}}$ and $\Theta \to 3H_{\sms{FRW}}$ in the FRW limit. We will use the correction from $\rho_{\sms{FRW}}$ to $\rho_{\rm eff}$ to represent the correction from $H_{\sms{FRW}}$ to $H_{\sms D}$ in Sec. IV.

\subsection{Evaluating the correction terms with large-scale structure simulations}

In order to connect the correction terms that we derived in Sec. IIA and IIB with presently available large-scale structure simulation codes, we begin with the conformal Newtonian gauge \cite{Mukhanov:1992,Ma:1995},
\begin{equation}
ds^2=a^2_{\sms{FRW}}(\eta)[-(1+2\Phi)d\eta^2+(1-2\Phi)dx^idx_i]~~.
\end{equation}
This describes a restricted class of general gauge-invariant cosmological perturbation theories \cite{Lifshitz:1946,Lifshitz:1963,Bardeen:1980,Kodama:1984}. We then identify this as the weak-field limit of our conformally flat metric [Eq. (\ref{ds})]. Here, $\Phi$ is the peculiar gravitational potential, $\eta$ is the conformal time, $d\eta={dt}/{a_{\sms{FRW}}(t)}$. Using the fact that for any given $\eta$, there is a corresponding $t$, we can make a mapping that $\alpha^2 \to 1+2\Phi$ and $\phi^4 \to 1-2\Phi$ to express the metric coefficients, $\alpha$ and $\phi$, in terms of $\Phi$.

By definition, the domain averaged expansion rate $H_{\sms D}$ is a constant within a local domain at a given redshift. Utilizing this, and the expression ${\dot{\phi}}/{\phi}=-{\dot\Phi}/{2(1-2\Phi)}$, $H_{\sms D}$ can be expressed as:
\begin{equation}
H_{\sms D} = \frac{1}{3}\langle-\alpha K\rangle_{\sms D} = \langle H_{\sms{FRW}}+2\frac{\dot{\phi}}{\phi} \rangle_{\sms D} =  H_{\sms{FRW}} + 2\langle\frac{\dot{\phi}}{\phi} \rangle_{\sms D} =  H_{\sms{FRW}} - \langle\frac{\dot\Phi}{1-2\Phi}\rangle_{\sms D} ~~.
\end{equation}
The determinant of $\gamma_{ij}$ in Eq. (\ref{gammaij}) now becomes, $\gamma = a^3_{\sms{FRW}}(t)\phi^6(\bm x,t) = a^3_{\sms{FRW}}(1-2\Phi)^{\frac{3}{2}}$, and by using the definition of the domain average in Eqs. (\ref{vd}) and (\ref{daverage}), $H_{\sms D}$ can be further written as:
\begin{equation}
H_{\sms D} = H_{\sms{FRW}} -\frac{\int_{\sms D}\frac{\dot\Phi}{1-2\Phi}\gamma\, d^3 x}{\int_{\sms D}\gamma\, d^3 x} = H_{\sms{FRW}} - \frac{\int_{\sms D}\dot\Phi(1-2\Phi)^\frac{1}{2}\, d^3 x}{\int_{\sms D}(1-2\Phi)^\frac{3}{2}\, d^3 x} ~~.
\end{equation}
Now, $H_{\sms D}$ is only dependent on $H_{\sms{FRW}}$ and $\Phi$. Both quantities can be easily extracted from large-scale structure simulations. The effective energy density $\rho_{\rm eff}$ can be calculated as:
\begin{equation}
\rho_{\rm eff} = \frac{3 H_{\sms D}^2}{8\pi G} = \frac{3}{8\pi G} H_{\sms{FRW}}^2 - \frac{3}{4\pi G}H_{\sms{FRW}}\frac{\int_{\sms D}\dot\Phi(1-2\Phi)^\frac{1}{2}\, d^3 x}{\int_{\sms D}(1-2\Phi)^\frac{3}{2}\, d^3 x}+\frac{3}{8\pi G}\left(\frac{\int_{\sms D}\dot\Phi(1-2\Phi)^\frac{1}{2}\, d^3 x}{\int_{\sms D}(1-2\Phi)^\frac{3}{2}\, d^3 x} \right)^2 ~~.
\end{equation}
The correction to the energy density, $(\rho_{\rm eff}-\rho_{\sms FRW})/\rho_{\sms FRW}$, can then be expressed as:
\begin{equation}
\label{rho_cor}
\frac{\rho_{\rm eff}-\rho_{\sms {FRW}}}{\rho_{\sms {FRW}}}=-\frac{2}{H_{\sms{FRW}}}\frac{\int_{\sms D}\dot\Phi(1-2\Phi)^\frac{1}{2}\, d^3 x}{\int_{\sms D}(1-2\Phi)^\frac{3}{2}\, d^3 x}+\frac{1}{H_{\sms{FRW}}^2}\left(\frac{\int_{\sms D}\dot\Phi(1-2\Phi)^\frac{1}{2}\, d^3 x}{\int_{\sms D}(1-2\Phi)^\frac{3}{2}\, d^3 x} \right)^2 ~~.
\end{equation}

The expression for the effective pressure $P_{\rm eff}$ in Eq. (\ref{2friedmann}) is somewhat cumbersome. Hence, we will analyze it term by term. Since for the current application, we only need to deal with a universe dominated by nonrelativistic matter, we can take  the matter pressure $P$ to be negligible. The first term in the expression of $P_{\rm eff}$ in Eq. (\ref{2friedmann}) can then be written as:
\begin{equation}
\frac{1}{3}\langle\alpha^2(\rho+3P)\rangle_{\sms D} = \frac{1}{3}\frac{\int_{\sms D}\rho(1+2\Phi)(1-2\Phi)^\frac{3}{2}\, d^3 x}{\int_{\sms D}(1-2\Phi)^\frac{3}{2}\, d^3 x} ~~.
\end{equation}
If we define the density fluctuation as $\delta\rho\equiv\rho-\rho_{\sms{FRW}}$ and use the fact $\rho_{\sms FRW} = {3 H_{\sms FRW}^2}/{8\pi G}$, we have:
\begin{equation}
\frac{1}{3}\langle\alpha^2(\rho+3P)\rangle_{\sms D}=\frac{1}{8\pi G}H_{\sms FRW}^2+\frac{1}{4\pi G}H_{\sms FRW}^2\frac{\int_{\sms D}\Phi(1-2\Phi)^\frac{3}{2}\, d^3 x}{\int_{\sms D}(1-2\Phi)^\frac{3}{2}\, d^3 x}+\frac{1}{3}\frac{\int_{\sms D}\delta\rho(1-4\Phi^2)(1-2\Phi)^\frac{1}{2}\, d^3 x}{\int_{\sms D}(1-2\Phi)^\frac{3}{2}\, d^3 x}  ~~.
\end{equation}
For the second term, we have $\dot{\alpha}={\dot{\Phi}}/{(1+2\Phi)^\frac{1}{2}}$, and $D_iD^i\alpha=\gamma^{ij}D_iD_j\alpha=3(a^2_{\sms{FRW}}(1-2\Phi))^{-1}\nabla^2(1+2\Phi)^\frac{1}{2}$. The entire second term can then be written as:
\begin{eqnarray}
\frac{1}{12\pi G}S_{\sms D}(\alpha)&=&-\frac{1}{4\pi G}\left(\langle\frac{\dot{\Phi}}{1+2\Phi}(H_{\sms{FRW}}-\frac{\dot\Phi}{1-2\Phi})\rangle_{\sms D}+\langle(1+2\Phi)^\frac{1}{2}\frac{\nabla^2(1+2\Phi)^\frac{1}{2}}{a^2_{\sms{FRW}}(1-2\Phi)}\rangle_{\sms D}\right) \nonumber \\
&=&-\frac{1}{4\pi G}\frac{1}{\int_{\sms D}(1-2\Phi)^\frac{3}{2}\, d^3 x}\left(\int_{\sms D}\frac{\dot{\Phi}}{1+2\Phi}(H_{\sms{FRW}}-\frac{\dot\Phi}{1-2\Phi})(1-2\Phi)^\frac{3}{2}\, d^3 x \right. \nonumber \\
& &\left. + \int_{\sms D}(1-4\Phi^2)^\frac{1}{2}a^{-2}_{\sms{FRW}}\nabla^2(1+2\Phi)^\frac{1}{2}\, d^3 x \right) ~~.
\end{eqnarray}
Using the fact that $\Phi \ll 1$ (or $\Phi/c^2 \ll 1$, if we explicitly denote the value of ``c") in the weak-field limit and the cosmic Poisson equation:
\begin{equation}
\label{poisson}
\nabla^2\Phi=4{\pi}Ga^2_{\sms{FRW}}\delta\rho ~~,
\end{equation}
we deduce that $a^{-2}_{\sms{FRW}}\nabla^2(1+2\Phi)^\frac{1}{2}=4{\pi}G\delta\rho$. So the second term can be further written as:
\begin{equation}
\frac{1}{12\pi G}S_{\sms D}(\alpha)=-\frac{1}{4\pi G}H_{\sms{FRW}}\frac{\int_{\sms D}\frac{\dot{\Phi}}{1+2\Phi}(1-2\Phi)^\frac{3}{2}\, d^3 x}{\int_{\sms D}(1-2\Phi)^\frac{3}{2}\, d^3 x}+\frac{1}{4\pi G}\frac{\int_{\sms D}\frac{\dot{\Phi}^2}{1+2\Phi}(1-2\Phi)^\frac{1}{2}\, d^3 x}{\int_{\sms D}(1-2\Phi)^\frac{3}{2}\, d^3 x}-\frac{\int_{\sms D}\delta\rho(1-4\Phi^2)^\frac{1}{2}\, d^3 x}{\int_{\sms D}(1-2\Phi)^\frac{3}{2}\, d^3 x} ~~.
\end{equation}
The third term is simply the $\rho_{\rm eff}$. The fourth term can be written as:
\begin{eqnarray}
-\frac{1}{18\pi G}\langle\Theta^2\rangle_{\sms D} &=& -\frac{1}{2\pi G} \frac{\int_{\sms D}(H_{\sms{FRW}} - \frac{\dot\Phi}{1-2\Phi})^2(1-2\Phi)^\frac{3}{2}\, d^3 x}{\int_{\sms D}(1-2\Phi)^\frac{3}{2}\, d^3 x} \nonumber \\
&=& -\frac{1}{2\pi G}H_{\sms{FRW}}^2 + \frac{1}{\pi G}H_{\sms{FRW}}\frac{\int_{\sms D}\dot\Phi(1-2\Phi)^\frac{1}{2}\, d^3 x}{\int_{\sms D}(1-2\Phi)^\frac{3}{2}\, d^3 x}-\frac{1}{2\pi G}\frac{\int_{\sms D}\frac{\dot\Phi^2}{(1-2\Phi)^\frac{1}{2}}\, d^3 x}{\int_{\sms D}(1-2\Phi)^\frac{3}{2}\, d^3 x} ~~.
\end{eqnarray}
The effective pressure $P_{\rm eff}$ can now be summarized as:
\begin{eqnarray}
\label{p_eff}
P_{\rm eff} &=& \frac{1}{4\pi G}H_{\sms FRW}^2\frac{\int_{\sms D}\Phi(1-2\Phi)^\frac{3}{2}\, d^3 x}{\int_{\sms D}(1-2\Phi)^\frac{3}{2}\, d^3 x} + \frac{1}{\pi G}H_{\sms{FRW}}\frac{\int_{\sms D}\dot\Phi\Phi(1-2\Phi)^\frac{1}{2}(1+2\Phi)^{- 1}\, d^3 x}{\int_{\sms D}(1-2\Phi)^\frac{3}{2}\, d^3 x} \nonumber \\
& & +\frac{1}{3}\frac{\int_{\sms D}\delta\rho(1-4\Phi^2)(1-2\Phi)^\frac{1}{2}\, d^3 x}{\int_{\sms D}(1-2\Phi)^\frac{3}{2}\, d^3 x} - \frac{\int_{\sms D}\delta\rho(1-4\Phi^2)^\frac{1}{2}\, d^3 x}{\int_{\sms D}(1-2\Phi)^\frac{3}{2}\, d^3 x} \nonumber \\
& & -\frac{1}{4\pi G}\frac{\int_{\sms D}\dot{\Phi}^2(1+6\Phi)(1+2\Phi)^{-1}(1-2\Phi)^{-\frac{1}{2}}\, d^3 x}{\int_{\sms D}(1-2\Phi)^\frac{3}{2}\, d^3 x} + \frac{3}{8\pi G}\left(\frac{\int_{\sms D}\dot\Phi(1-2\Phi)^\frac{1}{2}\, d^3 x}{\int_{\sms D}(1-2\Phi)^\frac{3}{2}\, d^3 x} \right)^2 ~~.
\end{eqnarray}
The equation of state parameter of the effective dark energy like term can be defined as $w_{\rm eff}={P_{\rm eff}}/({\rho_{\rm eff}-\rho_{\sms{FRW}}})$.

\section{Details of the numerical simulation}
In the present work, our goal is to estimate the magnitude of the deduced correction terms. To achieve this, we do a straightforward large-scale structure simulation in a standard FRW cosmology and post-process the simulation data to evaluate the correction terms at each given redshift. In subsequent works, we will evolve the simulation with the modified equations of motion in real time and evaluate the resulting cumulative effect.

The code we have adopted for the present numerical simulation is the N-body SPH code GADGET which was originally developed by Springel et al. \cite{Springel:2001}. The most current publicly available version, Gadget-2 \cite{Springel:2005}, is used for all of the simulations described in this paper.

We set up the initial condition for the simulation as follows: First, the initial linear matter power spectrum for one specific set of cosmological parameters is generated by standard CMB codes such as CMBFAST \cite{Seljak:1996}; We then use the associated power spectrum to generate a Gaussian random field with the Zel'dovich approximation \cite{Zeldovich:1970} utilizing such generator codes as the Grafic \cite{Bertschinger:2001} packages and the IC \cite{Sirko:2005} package; Finally, the file is converted into the Gadget-2 format to start the simulation within a periodic comoving box. Some authors \cite{Buchert:1997,Buchert:2000a,Buchert:2008,Paranjape:2008b} have argued that the application of a periodic boundary condition in numerical simulations is equivalent to forcing a standard FRW cosmology and thus eliminates any proposed corrections. However, for the tree algorithm in the Gadget-2 simulation \cite{Barnes:1986,Hernquist:1991,Springel:2001,Springel:2005} and the size of our domain, this is not necessarily the case as discussed below.

In order to calculate the new domain averaged terms, $\rho_{\rm eff}, P_{\rm eff}$ and $w_{\rm eff}$, as derived in Sec. II, we utilize the Gadget-2 output of particle masses and positions at each time slice. For the domain averaging procedure, we first divide the whole domain, i.e. the entire simulation box, into a fine spatial grid of size $L_d$. We then use a Cloud-in-Cell (CIC) \cite{Hockney:1981} method to assign the matter density $\rho$ to each zone within the grid. Finally, we use the cosmic Poisson equation [Eq. (\ref{poisson})] with the method of Successive Over Relaxation (SOR) \cite{Press:1992} to calculate the peculiar gravitational potential $\Phi$ for each zone. We note that the application of a periodic boundary condition in the simulation implies that the solution to Eq. (\ref{poisson}) for the whole grid is only unique up to an arbitrary constant, this is adequate for evolving the equation of motion because they only involve the gradient of $\Phi$. However, this periodic boundary condition approach does not determine the value of $\Phi$ uniquely as is needed to calculate the correction terms. Hence, we use a fixed comoving boundary assuming that outside of the simulation box, the matter density is exactly $\rho_{\sms{FRW}}$. In this way, $\Phi$ is uniquely determined by the matter density distribution within the simulation only. Once $\rho$ and $\Phi$ are obtained in this way, the domain averaged quantities defined in Sec. II can be calculated. The physical implication of this procedure is that by reducing the smoothing length from the Hubble scale to the resolution limit of the simulation, we can effectively evaluate the effect of the local inhomogeneities on the global cosmic expansion rate, which is always neglected in the standard FRW cosmology. This is the procedure that was proposed by Ellis \cite{Ellis:1984,Ellis:2008} and by Ellis and Stoeger \cite{Ellis:1987}. The accuracy of this approach is limited, however, by the number of particles in the simulation and the resolution of the spatial grid. In the next section, we examine the dependence of the effect on the size, resolution and number of particles in the simulation. We show that a converged result can be obtained.

\section{Results}

\begin{figure}[!htb]
\centering
		\includegraphics[angle=-90, width=8cm]{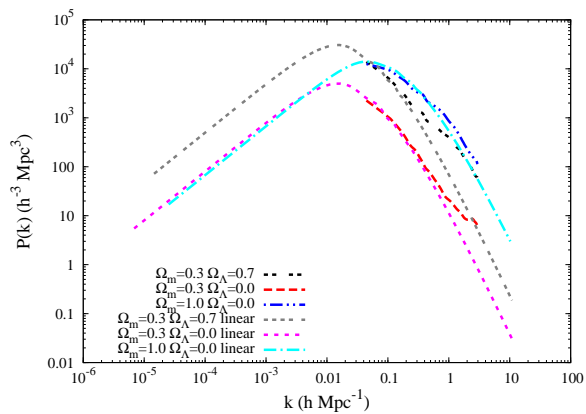}
	\caption{Matter power spectrum for three different cosmologies from both a linear theory and simulations.}
	\label{ps}
\end{figure}

\begin{figure}[!htb]
\centering
		\includegraphics[angle=-90, width=8cm]{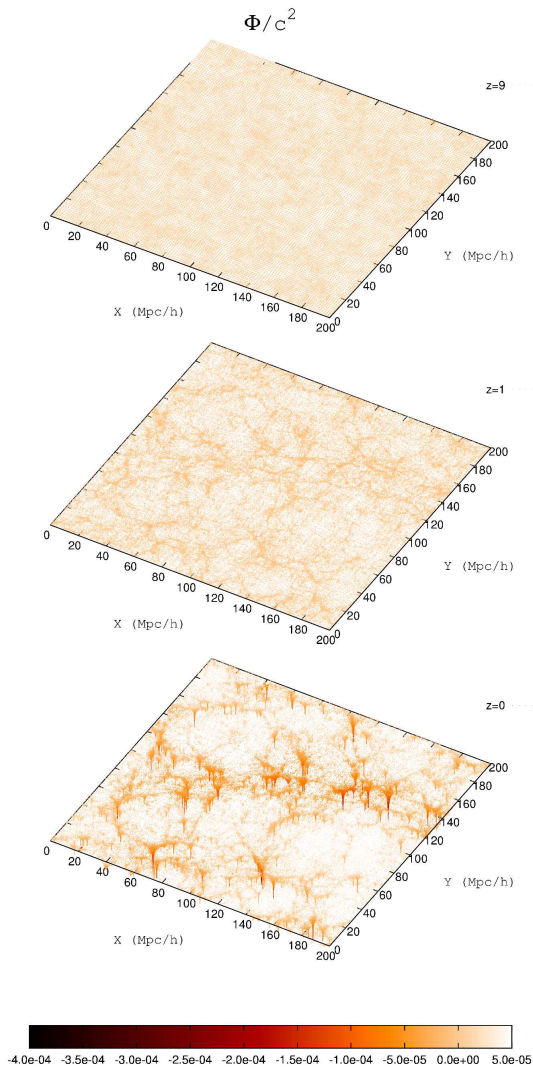}
	\caption{The growth of the peculiar gravitational potential $\Phi$ from redshift $z=9$ to $z=0$ on a comoving spatial slice in the X-Y plane in a three dimensional simulation. The value of $\Phi/c^2$ is drawn along the Z axis as indicated at the bottom of the figure.}
	\label{phi}
\end{figure}

As an illustration, Fig. \ref{ps} shows the matter power spectrum for three different cosmologies as labeled from our simulations at redshift z=0. On small scales, the simulations involve substantial nonlinear growth of structures that a linear theory cannot predict accurately. Fig. \ref{phi} shows the growth of the peculiar gravitational potential $\Phi$ during the structure formation epoch at three different redshifts in a flat, $\Omega_m = 1$ cosmology simulation. In all of these simulations, the upper limit of the absolute value of the metric perturbations, $\Phi/c^2$, is about $10^{-3}$ and the upper limit of the magnitude of the peculiar velocities is about $10^4$ km/s ($v/c \le 0.03$). Hence, the simulations indeed stay within the weak-field, slow-motion regime as assumed in the derivation in Sec. II.

Next, we study the proposed correction terms in detail using simulations in a flat, $\Omega_m=1$, matter-dominated cosmology. Table \ref{list} lists the various box sizes and numbers of particles we used for this special case.

\begin{table}[htdp]
\caption{List of the parameters of the simulations.}
\begin{center}
\begin{ruledtabular}
\begin{tabular}{cccccccccc}
Parameter & I & II & III & IV & V & VI & VII & VIII & IX \\
\hline
Box size (Mpc/h) $L$ & 100 & 100 & 200 & 200 & 400 & 400 & 800 & 800 & 1600 \\
Number of particles $N_p$ & $128^3$ & $256^3$ & $128^3$ & $256^3$ & $128^3$ & $256^3$ &$256^3$ & $512^3$ & $256^3$ \\
\end{tabular}
\end{ruledtabular}
\end{center}
\label{list}
\end{table}

\begin{figure}[!htb]
\centering
		\includegraphics[angle=-90, width=8cm]{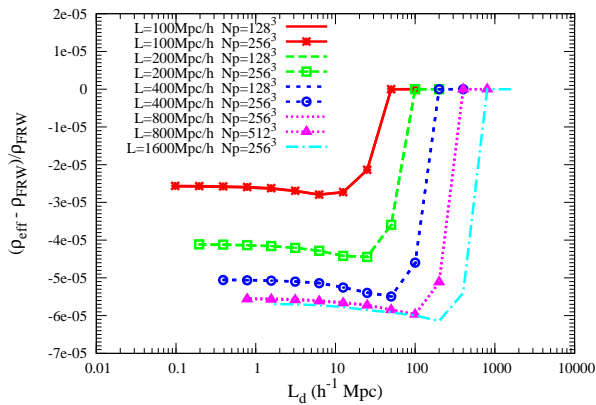}
	\caption{Correction to the energy density, $(\rho_{\rm eff}-\rho_{\sms FRW})/\rho_{\sms FRW}$ as a function of the smoothing length $L_d$ for various box sizes and numbers of particles. Note the convergence for the largest box sizes.}
	\label{rho_ld}
\end{figure}

\begin{figure}[!htb]
\centering
		\includegraphics[angle=-90, width=8cm]{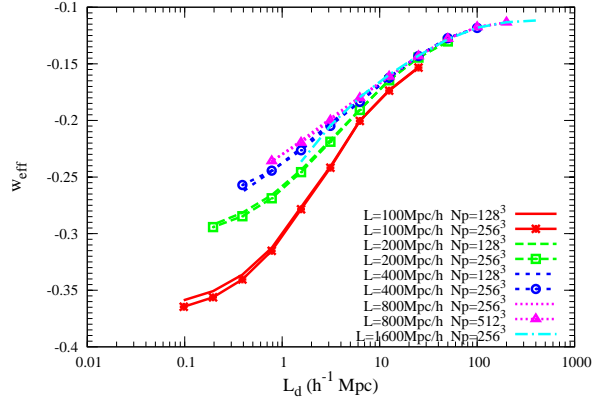}
	\caption{Effective EOS parameter $w_{\rm eff}$ as a function of the smoothing length $L_d$ for various box sizes and numbers of particles. Note the convergence for the largest box sizes.}
	\label{w_ld}
\end{figure}

Since the proposed correction terms arise from the domain averaging procedure, we wish to investigate their dependence on the smoothing length $L_d$, which is the individual zone size of the grid.\footnote{Note that $L_d$ is not the size of the whole domain, it is the smoothing length as described in \cite{Ellis:1984,Ellis:1987,Ellis:2008}.} Figs. \ref{rho_ld} and \ref{w_ld} plot the correction to the energy density $(\rho_{\rm eff}-\rho_{\sms FRW})/\rho_{\sms FRW}$ and the effective equation of state parameter $w_{\rm eff}$ as a function of $L_d$ at the current epoch. We have considered a set of $L_d$ values equal to the simulation box size $L$ divided by the powers of two, i.e. $L_d=L, L/2, L/4, \ldots$. We can see from Fig. \ref{rho_ld} that when $L_d$ is comparable to the box size $L$, i.e. $L_d=L, L/2$, the correction to the energy density is essentially negligible and we recover a standard FRW cosmology. This is because the local inhomogeneities in the individual zones are effectively smoothed out just as in the case when one simply uses the averaged energy density in a FRW cosmology. When $L_d$ further decreases from $L/4$, one can clearly see that the correction quickly grows and then converges to a negative value at about the $10^{-5}$ level. 

This can be explained by examining Eq. (\ref{rho_cor}) for $(\rho_{\rm eff}-\rho_{\sms FRW})/\rho_{\sms FRW}$. It is obvious that the second term in this expression is always positive. Since our simulations are in the weak-field, slow-motion regime, the magnitude of the second term is much smaller than that of the first term because it is second order, i.e. $\dot\Phi^2$. The sign of the first term is determined by the nature of the dominant volume weighted regions in the domain, i.e. collapsing overdense regions or expanding underdense regions. From Eq. (\ref{poisson}), one can see that collapsing or expanding regions have negative or positive $\dot\Phi$ terms, respectively. This leads to a positive or negative first term. From Fig. \ref{rho_ld}, the simulation results clearly indicate that the expanding underdense regions are the dominant volume weighted regions. This also explains why the correction remains invariant as $L_d$ further decreases. This is because once the resolution of the grid is fine enough to resolve the dominant regions, better resolution only slightly improves the accuracy of the correction. From Fig. \ref{rho_ld} we can also find that the correction is nearly independent of the number of particles in the simulations. This is due to the way the matter density is distributed on the grid as described in Sec. III. 

The magnitude of the correction grows with the size of the simulation box. This is because simulations with larger box sizes have less restriction on the nonlinear growth of structures from a FRW cosmology boundary and thus include larger structures. It is clearly shown in Fig. \ref{rho_ld} that the magnitude of the correction asymptotically converges to a value that is roughly represented by simulations with box sizes of 800 and 1600 Mpc/h. The most realistic asymptotic value is $-5.6 \times 10^{-5}$ at the best resolution of the grid in the simulations.

For the equation of state parameter $w_{\rm eff}$, we only plot the results with $L_d \le L/4$ in Fig. \ref{w_ld} because both ${P_{\rm eff}}$ and ${\rho_{\rm eff}-\rho_{\sms{FRW}}}$ are close to zero for $L_d=L, L/2$, so that large numerical errors are introduced into the calculated $w_{\rm eff}$ on these scales. It is shown in Fig. \ref{w_ld} that $w_{\rm eff}$ is always negative and has a magnitude of a few tenths for the plotted $L_d$ values. It decreases as $L_d$ decreases. For different simulations, $w_{\rm eff}$ increases with the box size and is almost independent of the number of particles in the simulations. Again, it asymptotically converges to a value given in the 800 and 1600 Mpc/h simulations. The most realistic value is about -0.24. Note that $w_{\rm eff}$ is negative because $\rho_{\rm eff}$ is less than $\rho_{\sms{FRW}}$, while ${P_{\rm eff}}$ is always positive. A cosmic acceleration requires a negative ${P_{\rm eff}}$. Therefore, even though $w_{\rm eff}<0$, no cosmic acceleration results. The sign of $w_{\rm eff}$ and ${P_{\rm eff}}$ can be explained by analyzing the expression for $P_{\rm eff}$, Eq. (\ref{p_eff}). Because our simulations are in the weak-field, slow-motion regime, all of the second order terms, e.g. $\Phi^2, \dot\Phi^2$ and $\Phi\dot\Phi$, are much smaller than the first order terms. Hence, the first and third terms in the expression of $P_{\rm eff}$ are the dominant terms. Since the expanding underdense regions are the dominant regions in all of our simulations, the first term is positive. The third term is also positive. This is because the collapsing overdense regions' densities are always weighted more than the underdense ones due to the $(1-2\Phi)^{\frac{1}{2}}$ factor no matter whether they are the dominant regions or not as long as $1-4\Phi^2>0$. Therefore the effective pressure $P_{\rm eff}$ is always positive. However, we will discuss one possible scenario in which one can have a large negative pressure term in Sec. V.

\begin{figure}[!htb]
\centering
		\includegraphics[angle=-90, width=8cm]{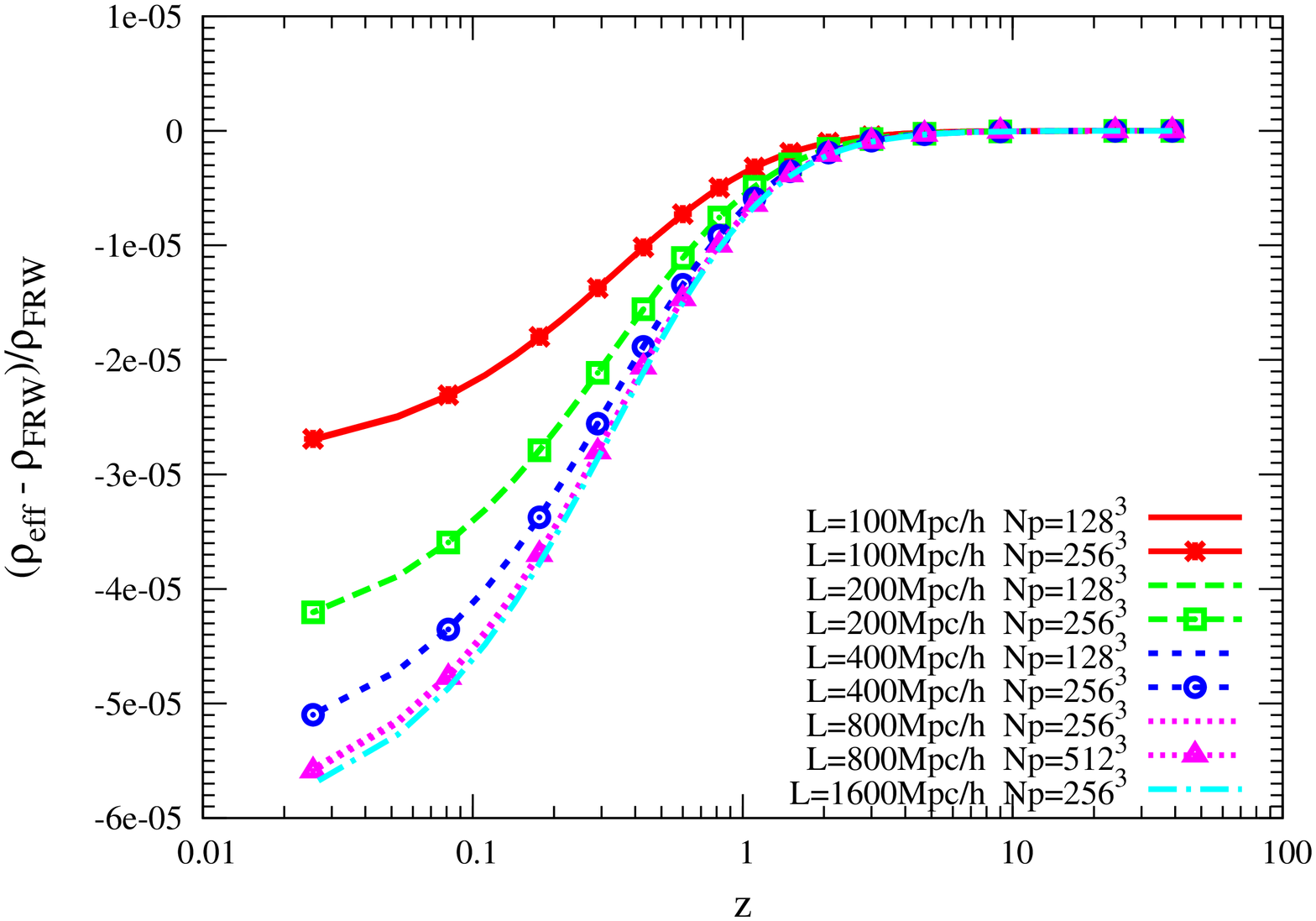}
	\caption{$(\rho_{\rm eff}-\rho_{\sms FRW})/\rho_{\sms FRW}$ as a function of redshift z for various box sizes and numbers of particles. Note the convergence for the largest box sizes.}
	\label{rho_z}
\end{figure}

\begin{figure}[!htb]
\centering
		\includegraphics[angle=-90, width=8cm]{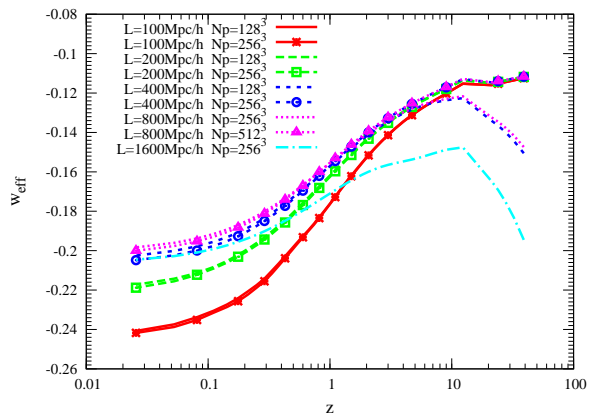}
	\caption{$w_{\rm eff}$ as a function of redshift z for various box sizes and numbers of particles. Note the convergence for the largest box sizes.}
	\label{w_z}
\end{figure}

As an illustration of the time evolution of the correction terms, Figs. \ref{rho_z} and \ref{w_z} show $(\rho_{\rm eff}-\rho_{\sms FRW})/\rho_{\sms FRW}$ and $w_{\rm eff}$, respectively, as a function of redshift z with $L_d= 3.125$ Mpc/h. From Fig. \ref{rho_z}, we can see that for all of the simulations, the correction to the energy density $(\rho_{\rm eff}-\rho_{\sms FRW})/\rho_{\sms FRW}$ grows from a negligible value at a redshift $z\sim40$ to a negative $10^{-5}$ level at the current epoch. The raw data show that it grows by about five orders of magnitude during this process. Starting from a redshift $z\sim2$, its magnitude quickly increases to the current level. The fact that it grows significantly during the structure formation process suggests it is the structure formation that leads to the correction to the energy density. In fact, this happens just as the dark energy becomes the dominant energy form in the Universe. This coincidence allows the possibility that there may yet be a possible connection between structure formation and the emergence of the dark energy. As before, the results represented by the simulations with box sizes of 800 and 1600 Mpc/h are the most realistic values.

Fig. \ref{w_z} shows that $w_{\rm eff}$ is always negative and its magnitude grows gradually from a redshift  $z\sim40$ to the current epoch in all of the simulations. The apparent deviations, especially for the $L=1600$ Mpc/h simulation, at high redshifts are due to the fact that both ${P_{\rm eff}}$ and ${\rho_{\rm eff}-\rho_{\sms{FRW}}}$ are very small at those redshifts and our simulations only have a limited number of particles and limited resolution, thus, some numerical errors are introduced into the values of $w_{\rm eff}$. This is not the case at low redshifts. For this reason and the trend we discussed before, we believe the result from the 800 Mpc/h with $512^3$ particles simulation is closest to the true value of $w_{\rm eff}$.

From the results we have presented in this section, we conclude that: 1) Numerical simulations can be used to calculate the proposed effect despite previous claims to the contrary \cite{Buchert:1997,Buchert:2000a,Buchert:2008,Paranjape:2008b}; 2) The correction to the energy density is negative and its magnitude is slightly larger than the previously claimed \cite{Vanderveld:2007,Behrend:2008} $10^{-5}$ level at the current epoch. It grows by about five orders of magnitude along with the structure formation process; 3) The effective pressure is always positive and its magnitude is roughly of the same order as that of ${\rho_{\rm eff}-\rho_{\sms{FRW}}}$. Hence, the effective equation of state parameter is negative and its value is about -0.2 to -0.3; and 4) The proposed correction terms in the weak-field, slow-motion limit are not able to drive the current cosmic acceleration for the simple reasons that their magnitudes are too small and the effective pressure is always positive.

\section{Summary}
We have used the ADM formalism to develop a practical scheme to calculate the proposed domain averaging effect in an inhomogeneous cosmology within the context of numerical large-scale structure simulations. We find that in the weak-field, slow-motion limit, the proposed effect implies a small correction to the global expansion rate of the Universe. Under this limit, our simulations are always dominated by the expanding underdense regions, hence the correction to the energy density is negative and the effective pressure is positive. However, whether this is still the case when strong-field gravity is included in a more general scenario needs to be further investigated. At least in the current investigation, the proposed effect cannot be the source of the current cosmic acceleration. We have done our analysis on each given redshift in standard FRW cosmology simulations, whether the cumulative effect can significantly change the expansion history of the Universe remains to be further studied. Nevertheless, the fact that this effect just begins to grow during the structure forming era allows the possibility that relativistic corrections from the development of the cosmic structure may have played a non-negligible role on the global dynamics of the Universe.

We wish to point out that one possible scenario exists in which one could have a large negative correction to the energy density and a large negative pressure. This occurs when the underdense regions are the overwhelmingly dominant regions and they expand very rapidly during the cosmic evolution. In this scenario, the second order terms in the expression for the effective pressure become the dominant terms and they can induce a large negative pressure. We have verified this prediction in simple toy models. Whether this can happen in a realistic cosmological model needs to be investigated with strong-field gravity and full GR. If this is the case, the proposed GR correction in an inhomogeneous cosmology model may yet be found to serve as one possible source of the current cosmic acceleration. Also, for a simple collapsing overdense region in a FRW cosmology box, we find that both the correction to the energy density and the effective pressure are positive. In this scenario, the cosmic expansion is effectively slowed. From these two cases, it is suggested that the effect of the proposed GR correction on the local expansion rate behaves like a positive ``feedback" on the structure formation. This aspect needs to be studied with a more realistic model and probably large-scale structure surveys. Unlike the conformal Newtonian gauge, the conformally flat model that we utilized in this paper can be applied to models beyond the weak-field, slow-motion limit. In future work, we will use this metric to investigate some of these aspects.

\section*{ACKNOWLEDGMENTS}
This work is partially supported by the U.S. Department of Energy under grant DE-FG02-95-ER40934 and by the Joint Institute for Nuclear Astrophysics (JINA) through NSF-PFC grant PHY08-22648. XZ wishes to thank Edmund Bertschinger for some useful discussions. XZ also wishes to thank In-Saeng Suh for useful discussions and for help in setting up the simulation codes on the Notre Dame high performance computing cluster.


\end{document}